\documentclass[12pt, draftclsnofoot, onecolumn]{IEEEtran}
\ifCLASSINFOpdf
\else
\fi
\usepackage{multirow}
\usepackage{flushend}
\usepackage{graphicx}
\usepackage{color}
\usepackage{epstopdf}
\usepackage[square,comma,sort&compress,numbers]{natbib}
\usepackage{amsmath}
\usepackage{amssymb}
\usepackage{algpseudocode,algorithm}
\newtheorem{lemma}{Lemma}
\DeclareMathSizes{10}{8}{7}{6}
\IEEEaftertitletext{\vspace{-2\baselineskip}}
\begin{document}
\title{On the Fundamental Limits of \\Random Non-orthogonal Multiple Access \\in Cellular Massive IoT}
\author{Mahyar~Shirvanimoghaddam,~\IEEEmembership{Member,~IEEE,}
                Massimo~Condoluci,~\IEEEmembership{Member,~IEEE,}
                 Mischa~Dohler,~\IEEEmembership{Fellow,~IEEE,}
                Sarah~J.~Johnson,~\IEEEmembership{Member,~IEEE}

\thanks{M. Shirvanimoghaddam is with the School of Electrical and Information Engineering, The University of Sydney, NSW, Australia (email: mahyar.shirvanimoghaddam@sydney.edu.au).

M. Condoluci and M. Dohler are with the Centre for Telecommunications Research, Department of Informatics, King's College London, UK (email: \{massimo.condoluci; mischa.dohler\}@kcl.ac.uk).

Sarah. J. Johnson is with School of Electrical Engineering and Computing, The University of Newcastle, NSW, Australia (e-mail: sarah.johnson@newcastle.edu.au).}}
\maketitle

\begin{abstract}
Machine-to-machine (M2M) constitutes the communication paradigm at the basis of Internet of Things (IoT) vision. M2M solutions allow billions of multi-role devices to communicate with each other or with the underlying data transport infrastructure without, or with minimal, human intervention.
Current solutions for wireless transmissions originally designed for human-based applications thus require a substantial shift to cope with the capacity issues in managing a huge amount of M2M devices. In this paper, we consider the multiple access techniques as promising solutions to support a large number of devices in cellular systems with limited radio resources. We focus on non-orthogonal multiple access (NOMA) where, with the aim to increase the channel efficiency, the devices share the same radio resources for their data transmission. This has been shown to provide optimal throughput from an information theoretic point of view. We consider a realistic system model and characterize the system performance in terms of throughput and energy efficiency in a NOMA scenario with a random packet arrival model, where we also derive the stability condition for the system to guarantee the performance.
\end{abstract}

\IEEEpeerreviewmaketitle

\section{Introduction}
\IEEEPARstart{I}{nternet} of Things (IoT) represents a major technology trend, which is revolutionizing the way we interact with our surrounding physical environment as our everyday physical objects will be transformed into information sources \cite{M2MMInternet,7123563}. The basic enabler for IoT is the massive connectivity between devices, e.g. sensors and actuators, and with the underlying data transport infrastructure without, or with limited, human interaction. Machine-to-machine (M2M) aims at providing this communication infrastructure for the emerging IoT applications and services in the near future \cite{TUbiq,7516570}. The most promising solution proposed for M2M communications is wireless cellular, e.g., GSM, GPRS, 3G, WiMAX, as well as Long Term Evolution (LTE) and LTE-Advanced (LTE-A), due to their excellent coverage, mobility and scalability support, good security features, and the availability of the infrastructure almost everywhere \cite{M2M_Ericsson2,7403840}. The focus of this study is on cellular M2M communications for massive IoT.

\subsection{Background and Motivations}
The third generation partnership project (3GPP) has already initiated several task groups to standardize several low-power solutions for emerging M2M communications, which are referred to as machine-type communications (MTC) in the 3GPP terminology. Such solutions include extended coverage GSM (EC-GSM), LTE for machine-type communication (LTE-M), and narrow band IoT (NB-IoT) \cite{M2M_Ericsson2,M2M_Nokia}. These standards have been proposed on top of existing cellular standards by exploiting new control and data channels to increase capacity per cell and power saving functionality to extend battery life \cite{M2M_Ericsson2}.

Moving into the future, major improvements in system performance will require a more substantial shift from current protocols and designs originally proposed for human based communications. The fact that the data channels are orthogonally allocated to the devices in current cellular systems makes it a potential bottleneck for future M2M applications, where a large number of devices want to communicate with the base station (BS) and there are not enough radio resources to be orthogonally allocated to the devices \cite{Mahyar_TWC}. We foresee that new multiple access (MA) techniques are essential for future cellular systems to enable multiple M2M devices to effectively share radio resources.

Multiple access techniques can be generally divided into orthogonal and non-orthogonal approaches. In orthogonal MA (OMA), radio resources are orthogonally divided between devices, where the signals from different devices are not overlapped with each other. Instances of OMA are time division multiple access (TDMA), frequency division multiple access (FDMA), orthogonal frequency division multiple access (OFDMA), and single carrier FDMA (SC-FDMA) \cite{China_NOMA}. OMA approaches have no ability to combat inter-cell interference; therefore careful cell planning and interference management techniques are required to solve the interference problem. \textit{Non-orthogonal multiple access (NOMA)} allows overlapping among the signals from different devices by exploiting power domain, code domain (such as code division multiple access), and inter-leaver pattern. In NOMA, signals from multiple users are superimposed in the power-domain and successive interference cancellation (SIC) is used at the BS to decode the messages. A comparison between NOMA and OMA has been provided in \cite{NOMA6,ZTE_Mahyar}. NOMA in general has been shown to achieve the multiuser capacity region both in the uplink and downlink and provides better performance than OMA \cite{China_NOMA}.

NOMA was studies in \cite{NOMA1,NOMA2} in cellular systems, where the potentials and challenges of NOMA were discussed in the downlink of cellular systems. The effect of user pairing for NOMA in cellular systems was studied in \cite{NOMA3}, and it was shown that a better rate performance is achieved when a near user is paired with a far user to perform donwlink NOMA. In \cite{NOMA_1}, several NOMA strategies, such as power domain NOMA and code domain NOMA, have been studied and it has been shown that NOMA is compatible with massive MIMO and OFDM. Authors in \cite{NOMA_2} studied the performance of NOMA in the downlink of cellular systems with randomly deployed users and showed that power allocation and data rate selection play important roles in the overall performance of the system.

NOMA was studied for the uplink cellular systems in \cite{NOMA4}, where a novel power control strategy was proposed and the outage probability was analyzed. Authors in \cite{NOMA_6} investigated the enhancement of cell-edge user throughput by using NOMA with SIC. An enhanced proportional fair based scheduling scheme for non-orthogonal multiplexed users with contiguous resource allocation to retain the SC-FDM property was also proposed in \cite{NOMA_7}. Fractional frequency reuse was also used for NOMA to further enhance the performance of cell-edge users. As shown in \cite{NOMA_5,NOMA_7},  nonorthogonal access with a SIC can significantly enhance the uplink system performance and improve cell-edge user throughput compared to the orthogonal access.  In \cite{NOMA_8}, a novel dynamic power allocation scheme was proposed for downlink and uplink NOMA scenarios with two users for more flexibly meeting various quality of service requirements. The proposed scheme was shown to strictly guarantee a performance gain over conventional orthogonal multiple access and offer more flexibility to realize different tradeoffs between the user fairness and system throughput.

Only few studies have considered NOMA for massive number of devices in IoT. In \cite{NOMA_3}, a MIMO-NOMA strategy was designed for IoT, where only two users are paired to satisfy the service requirements of one of them while the other user is served opportunistically. Authors in \cite{NOMA_4} proposed a new type of non-orthogonal multiple access scheme called multi-user shared access, which adopts a grant-free access strategy to simplify the access procedure significantly and utilizes advanced code domain non-orthogonal
complex spreading to accommodate massive number of users in the same radio resources. The Energy efficiency of NOMA for uplink cellular M2M system was studied in \cite{NOMA_5}, and it was shown that transmitting with minimum rate and full time is optimal in terms of energy. These studies show that NOMA can support massive connectivity and grant-free access for massive IoT which are driving forces to study NOMA for IoT applications and services \cite{mag_Mahyar}. All of these studies rely on the channel assumption at both the transmitter and receiver side, where optimal resource and power allocation strategies can be designed. However, in real scenarios these assumptions are not valid due to a large number of devices in massive IoT. There is also a lack of system wide performance evaluation of cellular uplink systems for massive IoT with randomly deployed users and random traffic. In this paper, we consider random NOMA and evaluate the systems performance and derive the necessary condition for the stability of the system and characterize the maximum system throughput under the stability condition.

\subsection{Contributions and Paper Organization}
In uplink NOMA multiple devices simultaneously perform transmission in a shared radio resource; therefore, their transmissions are overlapping \cite{China_NOMA,higuchi_NOMA,ZTE_Mahyar}. NOMA has been already studied for multiple access in both uplink and downlink of wireless cellular networks, where the number of devices is usually assumed to be very small, e.g., 2 or 3 users, and the channel state information is available to optimize the transmit power. However, these assumptions are not valid anymore in massive IoT. We propose a NOMA-based multiple access strategy for massive IoT with random packet arrivals. In the proposed approach, referred to as random NOMA, each device which has data to transmit randomly chooses a sub-band and encodes its message along with its terminal identity (ID) and sends the encoded packet over the selected sub-band. As multiple devices may have selected the same sub-band, their transmissions interfere with each other. The base station can then perform successive interference cancellation (SIC) to decode the devices' messages, which is enabled by using rateless codes by the devices.

We derive the necessary condition for the stability of the system under the proposed random NOMA strategy. We find the maximum arrival rate for a system with an initial backlog such that the number of devices which are attempting to transmit to the BS in the next time slot does not increase in time. We consider two scenarios, 1) with no delay constraint and 2) with strict delay constraint to derive the system stability condition and characterize the maximum supportable arrival rate. We find the optimal resource allocation strategy, where the optimal number of sub-bands is found for a given available bandwidth to maximize the throughput of random NOMA. We showed that only a few subband, i.e. 2 or 3 subbands, are enough to maximize the system throughput without any delay constraint. On the other hand, when a strict delay constraint is imposed to the system, higher throughput is achieved when the whole bandwidth is used by a single subband.

The remainder of the paper is organized as follows. The system model is presented in Section II. The random NOMA strategy is proposed in Section III. The proposed NOMA strategy is analyzed in Section IV, where we derive the stability condition for the system and characterize the maximum packet arrival rate at the base station. System parameters are optimized in Section V. In Section VI, some practical considerations of massive NOMA in massive IoT are presented. Finally, Section VII concludes the paper.

\section{System Model}
We consider a single-cell wireless network consisting of one BS located at the centre and MTC devices are randomly distributed around the BS in an angular region with inner and outer radii $R_\mathrm{i}$ and $R_\mathrm{o}$ according to a homogeneous Poisson point process (PPP). As shown in  \cite[Appendix A]{PPP2} in a spatial PPP,  a device is located at coordinate $(r, \theta)$ with probability $p_{r,\theta}(r,\theta)$ defined as follows:
\begin{align}
p_{r,\theta}(r,\theta)=\frac{r}{\pi\left(R^2_o-R^2_i\right)},~~ R_i\le r\le R_o,~0\le\theta\le2\pi.
\end{align}
This model has been widely used in the literature and is the baseline assumption for many cellular system studies \cite{PPP1,3gpp_36212}. Table \ref{tab1} summarizes the notations commonly used in this paper.

\begin{table}[t]
\caption{Notation Summary}
\label{tab1}
\centering
\scriptsize
\begin{tabular}{|p{1cm}|p{7cm}|}
\hline
\textbf{Notation}&\textbf{Description}\\
\hline
$R_{\mathrm{i}}$&Inner cell radius\\
\hline
$R_{\mathrm{o}}$&Outer cell radius\\
\hline
$N_s$& Number of frequency sub-bands\\
\hline
$W$& Total available bandwidth (Hz)\\
\hline
$W_s$& The bandwidth of a frequency sub-band (Hz)\\
\hline
$\alpha$& path loss exponent\\
\hline
$g_i$& channel gain of the $i^{th}$ device to the BS\\
\hline
$r_i$& distance between the $i^{th}$ device and the BS\\
\hline
$\mu_{\mathrm{ref}}$& reference SNR\\
\hline
$P_{t}$& transmit power of MTC device\\
\hline
$P_{\max}$& maximum transmit power at an MTC device\\
\hline
$\mu_r$& Received SNR at the BS from a device at distance $r$\\
\hline
$G$& antenna gain\\
\hline
$\chi$& large scale shadowing gain\\
\hline
$h$& small scale fading gain\\
\hline
$\lambda$& New packet arrival rate at the BS\\
\hline
$L$& MTC message size (bits)\\
\hline
$M_s$& Number of available seeds\\
\hline
$T(n)$& duration of a time slot when $n$ devices are transmitting\\
\hline
$q(c,n)$& probability that the maximum number of devices over all the sub-bands is $c$ when the number of active devices in $n$.\\
\hline
$t(k)$& time slot duration of a sub-band containing $k$ devices\\
\hline
$P_c$& collision probability\\
\hline
\end{tabular}
\end{table}
\normalsize

For simplicity, we assume the BS and MTC devices are equipped each with a single antenna. We assume that radio resources are divided into $N_{\mathrm{s}}$ frequency sub-bands each with bandwidth $W_s=W/N_s$, where $W$ is the total available bandwidth. Following \cite{FunThroM2M,PowerEfficient,ZTE_Mahyar}, the channel between each MTC device and the BS is modeled by path loss, shadowing and small scale fading. The received power at the BS from an MTC device located at distance $r$ with transmit power $P_{\mathrm{t}}$ is given by:
\begin{align}
P_{\mathrm{r}}=P_{\mathrm{t}}\chi hG r^{-\alpha},
\end{align}
where $\alpha$ is the path loss exponent, $\chi$ is the large scale shadowing gain, $h$ is the small scale fading gain, and $G$ is the antenna gain. Similar to \cite{PowerEfficient}, we introduce the term reference signal-to-noise ratio (SNR), $\mu_{\mathrm{ref}}$, which is defined as the average received SNR from a device transmitting at maximum power $P_{\max}$ over the whole bandwidth $W$ located at the cell edge, i.e. at distance $R_\mathrm{o}$. The received SNR can then be expressed as follows \cite{PowerEfficient}:
\begin{align}
\mu_{r}=\frac{P_{\mathrm{t}}}{P_{\max}}\mu_{\mathrm{ref}}\chi h \left(\frac{r}{R_{\mathrm{o}}}\right)^{-\alpha}.
\label{eq:refsnr}
\end{align}
As information symbols might be transmitted over a smaller bandwidth $W_{\mathrm{s}}$, the effective noise power will be reduced by a factor $W/W_{\mathrm{s}}$. Therefore, the received SNR from an MTC device located at distance $r$ from the BS and transmitting over bandwidth $W_{\mathrm{s}}$ can be expressed as follows:
\begin{align}
\mu_r=\frac{P_{\mathrm{t}}}{P_{\max}}\frac{W}{W_{\mathrm{s}}}\mu_{\mathrm{ref}} \chi h \left(\frac{r}{R_{\mathrm{o}}}\right)^{-\alpha}.
\end{align}

We assume that the channel gain $\chi h (r/R_{\mathrm{o}})^{-\alpha}$ varies very slowly in time and is known at the MTC device. This is particularly advantageous for many fixed-location MTC applications as the device location is usually fixed and the MTC device can obtain accurate channel information in a timely manner. Moreover, the devices can perform the channel estimation by using regular pilot signals transmitted by the BS. This assumption will significantly  reduce the complexity at the BS as it does not need to estimate the channel to a very large number of MTC devices. With the channel estimation at the devices, we can also assume that the devices can mitigate the multi-path effect. There are two major assumptions which enables this, which are  \emph{Channel reciprocity}, the impulse responses of the forward link channel and the backward link channel are assumed to be identical, and \emph{Channel stationarity}, the channel impulse responses (CIRs) are assumed to be stationary for at least one probing-and-transmitting cycle. These two assumptions can be well satisfied for fixed-location MTC applications \cite{IoT_TRDMA}.
\begin{figure*}[t]
\centering
\includegraphics[width=1\columnwidth]{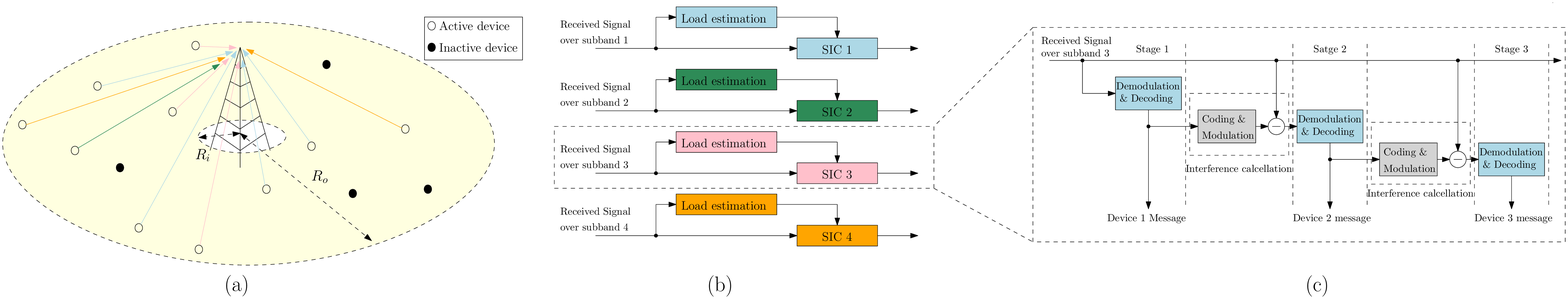}
  \caption{(a) Cellular MTC, where Each MTC device randomly chooses a sub-band for its transmission. (b) Load estimation at the base station for each subband. (c) successive interference cancellation and multi-level decoding at the base station over each subband.}
  \label{fig:SystemModel}
\end{figure*}

There are several approaches to mitigate the multi-path effect, such as time reversal (TR) strategy \cite{GreenTRDMA,IoT_TRDMA}, which has been recently studied for M2M communications as an effective strategy to focus signal waves in both time and space domains; thus improving system throughput and energy efficiency. In TR, after estimating the channel by the transmitter, it simply time reverses (and conjugated, if the signal is complex valued) the channel response waveform and then transmits it back through the same channel to the receiver. By utilizing channel reciprocity, the re-emitted TR waves can retrace the incoming paths, ending up with a constructive sum of signals of all the paths at the intended location and a ``spiky'' signal-power distribution over the space, as commonly referred to as \emph{spatial focusing effect}.  Therefore, we can ignore small-scale fading and shadowing, thus the channel gain and the transmit power is mainly characterized by the distance of the MTC devices to the BS.

Unless otherwise specified in the paper, each MTC device is assumed to control its transmit power using the channel information, such that the received SNR at the BS is $\mu_{0}$. Therefore, the transmit power required for an MTC device located at distance $r$ from the BS to achieve a received SNR $\mu_0$ over bandwidth $W_\mathrm{s}$ is given by:
\begin{align}
P_{\mathrm{t}}=P_{\max}\frac{\mu_0}{\chi h\mu_{\mathrm{ref}}}\left(\frac{r}{R_\mathrm{o}}\right)^{\alpha}.
\label{eq:transmitpower}
\end{align}

Similar to \cite{FunThroM2M,PowerEfficient}, we assume that the packet arrival rate at the BS follows a Poisson distribution with mean $\lambda$ packets per second. That is the number of packet transmission requests in a time interval of duration $t$ is given by $\mathrm{Poiss}(\lambda t)$. Each MTC device is assumed to have a message of length $L$ bits, including the device unique ID. Moreover, we consider slotted transmission and each device requests for a transmission only at the beginning of a time slot.

\section{The Proposed Random NOMA Strategy for M2M Communications}
In the proposed random NOMA strategy, the devices use the same radio resources for their transmissions. That is a device randomly chooses a sub-band for its data transmission and sends its data through the selected sub-band. The details of the proposed random NOMA strategy are given below:
\begin{itemize}
\item[1.] At the beginning of a time slot, the BS broadcasts a pilot signal over each sub-band.
\item[2.] Each MTC device which has data to transmit will randomly select a subband and listen to the pilot signal transmitted by the BS over that subband (Fig. \ref{fig:SystemModel}-a). It then estimates the channel over that subband. It also  randomly selects a seed for its random number generator from a set of $M_s$ available seeds.
\item[3.] Each active device attaches its unique ID to its message and encodes it using a Raptor codes constructed from the selected seed and transmits the codeword over the selected sub-band.
\item[4.] The BS performs load estimation (Fig. \ref{fig:SystemModel}-b) and successive interference cancellation (SIC) over each subband to recover the message of each active device (Fig. \ref{fig:SystemModel}-c). The SIC order is such that the BS starts the decoding with the first seed and remove the interference of it and then continues to the second seed and so on.
\end{itemize}

We assume that the BS broadcasts pilot signals over all the sub-bands. Using these pilot signals, each device will estimate its channel to the BS over the randomly selected subband. It is important to note that each device can also choose a sub-band which has the highest channel gain for its data transmission to reduce the energy consumption, however, this requires the devices to estimate the channels over all the subbands which might be only feasible when the number of subbands is small. As the devices perform power control such that their received power at the BS is the same, the BS can effectively estimate the number of devices over each sub-band by calculating the received power as it would be proportional to the number of devices. For further details on load estimation algorithms, please refer to \cite{Mahyar_SPM_Raptor}.

\begin{figure}[t]
  \centering
  \includegraphics[width=0.4\columnwidth]{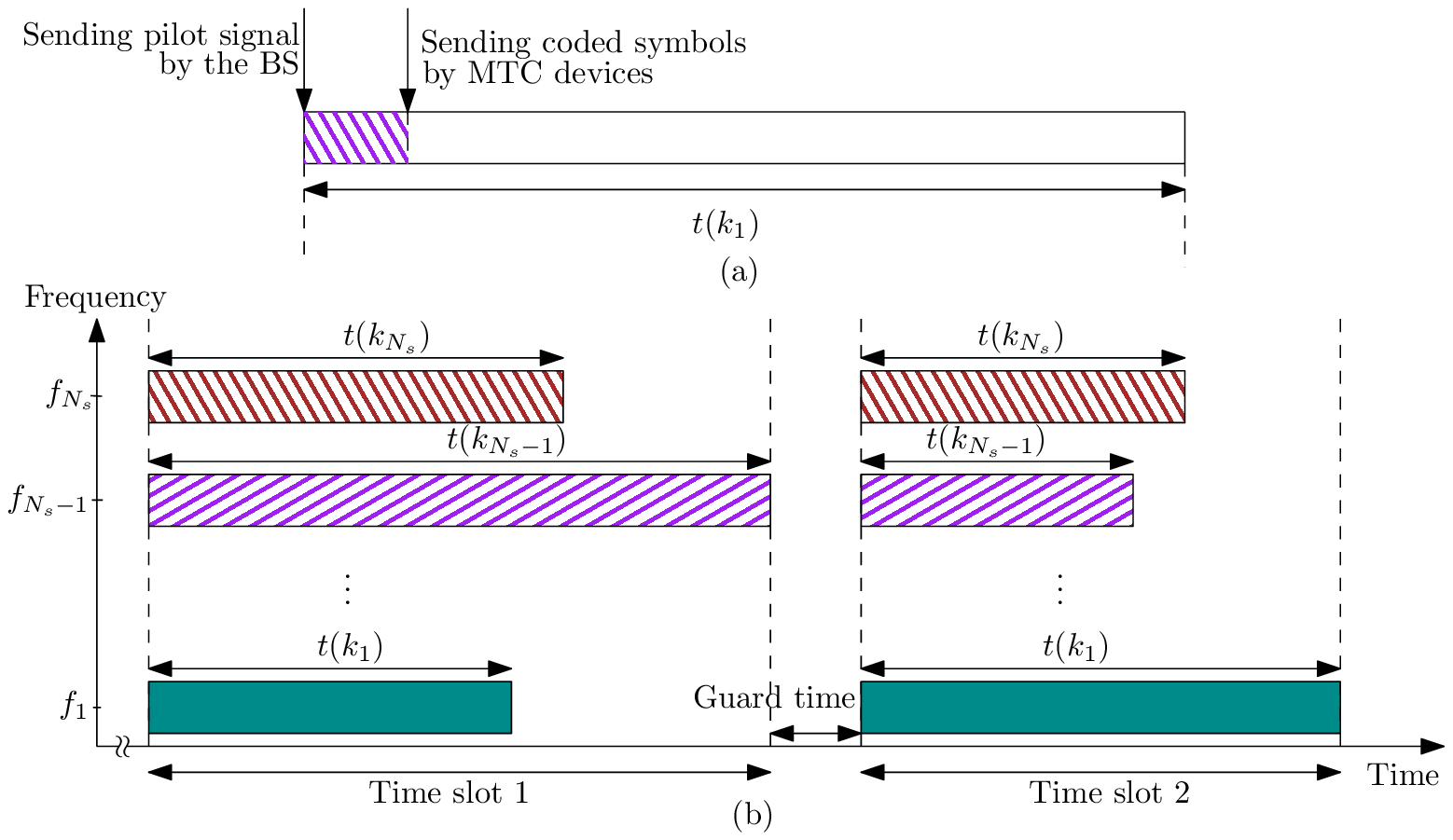}
  \vskip-1.5ex
  \caption{Time slot duration in the proposed random NOMA strategy (a) over a single sub-band (b) in two consecutive time slots.}
  \label{fig:subband}
\end{figure}

Due to the random number of active devices in each sub-band, the achievable rate over each subband is not fixed and depends on the number of devices, which is random. This means that the number of coded symbols that need to be transmitted over each sub-band is random. Fig. \ref{fig:subband}-b shows the length of each sub-band in two consecutive time slots. It is important to note that the duration of each time slot will be mainly determined by the sub-band with the highest number of active devices as its maximum achievable rate would be lower than the rest. This means that a fixed-rate code cannot be used in all time instances. Instead we propose to use Raptor codes \cite{Raptor}, which can generate as many coded symbols as required by the BS \cite{Mahyar_SPM_Raptor}. The code structure is random and is usually represented by a degree distribution function. Similar to LDPC codes, the code can be represented by a bipartite graph, as each coded symbol in a Raptor code is the XOR of a set of randomly selected information symbols.  The number of information symbols which are XORed to generate a coded symbol is random and is described by a degree distribution function. To implement a Raptor code, a pseudo random number generator is used in both the transmitter and receiver, and they need to use the same seed to start generating random numbers. This way the receiver can construct the same bipartite graph as that in the transmitter, therefore can perform belief propagation and decode the message. It is important to note that when more than one device selects the same seed and transmits over the same sub-band, they will be transmitting using exactly the same code structure; thus the BS cannot differentiate between them as there is no structural difference between the received codewords. We call this event a \emph{collision}.

To reduce the collision, we assume that each active device randomly selects its seed from a pool of seeds of size $M_s$. When performing successive interference cancellation over a particular subband by the BS, it initiates the decoding with the first seed. If it cannot decode any message, it changes the seed and reattempts the decoding. In this way, those devices which have selected non-collided seeds will be decoded and later will be acknowledged by the BS. This means that the BS may need to reattempt the decoding $M_s$ times per sub-band. It is also possible to assume that each device always use a unique seed which is determined in advance according to its unique ID. This way we can completely eliminate the collision. However, as the set of active devices is random and the BS does not know which devices are active, it needs to consider all possible seeds to perform SIC. But this is impractical as the number of devices, and accordingly unique seeds, are very large in massive MTC, therefore it is impossible to try every possible seeds. Using a pool of seeds with limited size $M_s$ helps to significantly reduce the complexity at the BS; while the collision rate can be well controlled by changing the pool size according to the traffic load.

\section{Analysis of the proposed scheme}
In this section, we analyze the proposed random NOMA strategy in terms of throughput and derive the stability condition, where we find the maximum arrival rate such that the system remains stable. For this aim, we first need to characterize the time slot duration, as it is random due to the random number of devices over the sub-bands, and then find the collision probability in order to find the number of devices which reattempt their transmissions in the next time slot due to collision.  It is important to note that for the purposes of our analysis in this paper, we ignore the pilot signal transmission time and guard time.

\subsection{Time slot duration}
\begin{lemma}
\label{lemma1}
Let $k_i$ denote the number of active devices which have selected the $i^{th}$ subband and $c=\max_{1\le i\le N_s}\{k_i\}$. The probability mass function of $c$, denoted by $q(c,n)$, when the number of active devices is $n$ and the number of subbands is $N_s$, can be approximated by
\begin{align}
q(c,n)\approx\frac{N_s\varphi\left(\frac{c-\frac{n}{N_s}}{\sigma_s}\right)\left[1-\Phi\left(\frac{c-\frac{n}{N_s}}{\sigma_s}\right)\right]^{N_s-1}}{\sigma_s},
\label{eq:finalAppqc}
\end{align}
where $\varphi(x)=\frac{1}{\sqrt{2\pi}}e^{-\frac{x^2}{2}}$, $\sigma_s=\sqrt{\frac{n}{N_s}(1-\frac{1}{N_s})}$, and $\Phi(x)=\int_{x}^{\infty}\varphi(t)dt$ is the cumulative distribution function ($\mathrm{c.d.f.}$) of the standard normal distribution.
\end{lemma}
\begin{IEEEproof}
See Appendix A for the proof.
\end{IEEEproof}
The duration of a time slot is determined by the sub-band with the highest number of active devices transmitting in it. The time slot duration can then be calculated from Shannon's capacity formula as follows:
\begin{align}
t(c)=\frac{L}{W_s\log_2\left(1+\frac{\mu}{1+(c-1)\mu}\right)}, ~~c=\max_{i} k_i,
\label{eq:timeduration}
\end{align}
where $\mu=\frac{W}{W_s}\mu_0=N_s\mu_0$, as each device transmits over a sub-band with bandwidth $W_s$ rather than $W$. The average time slot duration can then be calculated as follows:
\begin{align}
\overline{T}(n)=\sum_{c=\lceil n/N_s\rceil}^{n}q(c,n) \frac{L}{W_s \log_2\left(\frac{1+c\mu}{1+(c-1)\mu}\right)},
\label{eq:AndTimeDuration}
\end{align}
where $\lceil .\rceil$ is the ceil operator.

in practice the data of each device can be decoded at the BS when the SNR is larger than a threshold. We have actually incorporated that in our analysis, by calculating the achievable common rate for the devices over each subband, which is equivalent to a threshold SNR. When the code rate is less than that common rate, the devices can be decoded. By using rateless codes, we make sure that the devices continue their transmission until the base station receives a sufficient number of coded symbols, such that the realized rate is less than the achievable rate, therefore successfully decodes the message.

It is important to note that here, we have assumed that all the active devices have messages of the same length. The case where the devices have different message lengths can be easily analyzed using (\ref{eq:timeduration}) and (\ref{eq:AndTimeDuration}), where the message length for each device belongs to a set of message lengths $\{L_1, L_2, \cdots, L_{N_{\ell}}\}$, where $L_1<L_2<\cdots<L_{N_{\ell}}$. We can also consider that a set of subbands are allocated for each message length, that is the devices with a particular message length are only select their subbands from a particular set of subbands. More specifically, when there are $N_s$ subbands and $N_{\ell}$ possible message lengths, the subbands will be divided into $\lfloor N_s/N_{\ell}\rfloor$ groups, and the devices with message length $L_i$ will only choose their subbands from the $i^{th}$ group of subbands. The time slot duration with some approximations can then be found by using (\ref{eq:AndTimeDuration}) and replacing $L$ with $\bar{L}=\frac{\sum_{i=1}^{N_{\ell}} L_i}{N_{\ell}}$. This is valid as we have assumed that the devices are randomly distributed in the cell and each device has a particular message length with probability $1/N_{\ell}$.

It is also important to note that the devices in massive IoT applications usually require small data rate and can tolerate delay. This has been emphasized in \cite{M2M_Ericsson2,M2M_Nokia,M2M_America}. In the proposed scheme, we have assumed that the devices require the same data rate, or at least those devices which are transmitting at the same subband have the same packet length and also same data rate. In (\ref{eq:AndTimeDuration}), we have considered the common rate, which is the minimum data rate achieved by all the devices, and found the average time slot duration. To maximize the achievable rate for each device, the interference cancellation order must be carefully chosen, but we did not consider this case, as the base station does not know the channels and treat all the devices similarly. It is obvious that higher data rates for the devices can be achieved when choosing appropriate SIC order, but this requires complex optimization which requires channel knowledge of all the devices at the BS. We therefore assume that the common rate is achieved by all the device, and the common rate has been defined as the minimum rate achieved by all the device. This way the SIC order is not relevant. This decoding strategy is indeed suboptimal, however it is enough for massive IoT as the devices do not require high data rates and can usually tolerate delay.

\subsection{Collision probability}

It is clear that the BS cannot detect the devices which have selected the same seed and are transmitting at the same sub-band. This is because there is no structural difference between the transmitted codewords from two devices which have selected the same seed and are transmitting over the same sub-band. As the number of sub-bands is $N_s$, each device can randomly select a sub-band and a seed among $M_s\times N_s$ different options. The collision probability, defined as the probability that a given device selects a subband and seed which have already been selected by one or more other devices, can be approximated as follows when the number of active devices is $n$:
\begin{align}
\mathrm{P_c}(n)\approx1-\left(1-\frac{1}{M_s N_s}\right)^{n-1},
\label{eq:collisionProb}
\end{align}
which follows from the fact that a given device selects a specific preamble and seed with probability $1/(M_sN_s)$, and it is not in collision if other devices select different preambles and seeds, which happens with probability $(1-1/(M_sN_s))^{n-1}$. As the given device can select any of the $M_sN_s$ configurations of preambles and seeds, the probability that a given device is not in collision is simply $M_sN_s(1/(M_sN_s))(1-1/(M_sN_s))^{n-1}$. The probability of collision is then easily derived as (\ref{eq:collisionProb}).

We determine the minimum number of seeds required for the system to have a collision probability at most $\mathrm{P_c}(n)$ as follows:
\begin{align}
M_s\ge \frac{1}{N_s \left(1-\left(1-\mathrm{P_c}(n)\right)^{\frac{1}{n-1}}\right)}.
\end{align}
This can be further simplified for $\mathrm{P_c}(n)\to 0$ as follows \cite{PowerEfficient}:
\begin{align}
M_s\ge \frac{n-1}{N_s \mathrm{P_c}(n)}.
\end{align}

\subsection{Stability Condition without Delay Constraint}
We define the stability condition such that in the steady state, the number of active devices in the next time slot, including the collided devices in the previous time slot and the newly generated packets, is not larger than the number of devices in the previous time slot. This way we make sure that the system can support all the active devices in each time slot and the number of active devices does not increase in time; otherwise the system will be quickly saturated. As the number of devices in each time slot is a random variable, we can define two stability conditions as follows.
\subsubsection{Weak Stability Condition}
The first stability condition, which we refer to as the \emph{weak stability condition}, is defined based on the steady state average number of devices that can be supported by a system. For a system with an initial backlog, $n_1$, the weak stability condition is defined as follows:
\begin{align}
\mathbb{E}[n_{i+1}|n_i]\le n_i,~\mathrm{for}~ i\ge 1,
\label{weakstabcond}
\end{align}
where $n_i$ denote the number of active devices in the $i^{th}$ time slot, which incudes the devices which have been active in the $i^{th}$ time slot and those devices which have collided in the previous time slot and reattempt their transmission in the $i^{th}$ time slot. The following lemma gives the maximum arrival rate under the weak stability condition.
\begin{lemma}
Under the weak stability condition, the maximum arrival rate that can be supported by the BS is given by:
\begin{align}
\lambda^{\mathrm{(weak)}}_{\max}=\max_{n}\{\lambda^{\mathrm{(weak)}}_{\max}(n)\},
\label{weakfinalcond}
\end{align}
where
\begin{align}
\lambda^{\mathrm{(weak)}}_{\max}(n)=\frac{n(1-\mathrm{P_c}(n))}{\overline{T}(n)}.
\label{eq:weaknodelay}
\end{align}
\end{lemma}
\begin{IEEEproof}
See Appendix B.
\end{IEEEproof}

\subsubsection{Strong Stability Condition}
The second stability condition, referred to as the \emph{strong stability condition}, takes into account the random behavior of the system and is given by:
\begin{align}
\mathbb{P}\left[n_{i+1}>n_i\right]\le \epsilon,
\end{align}
where $\epsilon>0$ is the target probability of the system stability, which is a system design parameter. The following lemma gives the maximum arrival rate under the strong stability condition.
\begin{lemma}
The maximum arrival rate under the strong stability condition is then given by
\begin{align}
\lambda^{\mathrm{(strong)}}_{\max}=\displaystyle\max_{n}\left\{\lambda^{\mathrm{(strong)}}_{\max}(n)\right\},
\label{eq:strongwodelay}
\end{align}
where
\begin{align}
\lambda^{\mathrm{(strong)}}_{\max}(n)=\frac{1+2n\ell_\epsilon(1-\mathrm{P_c}(n))-\sqrt{1+4n\ell_\epsilon(1-\mathrm{P_c}(n))^2}}{2\ell_\epsilon\overline{T}(n)},
\label{eq:strongcloseform}
\end{align}
and
\begin{align}
\ell_\epsilon:=\left(\Phi^{-1}(\epsilon)\right)^{-2}.
\label{eq:ellepsilon}
\end{align}

\end{lemma}

\begin{IEEEproof}
See Appendix C.
\end{IEEEproof}

\subsection{Stability Condition with QoS Guarantee}
In the proposed scheme, the time slot duration is not fixed and depends on the number of devices which have selected each subband.  However, time slot duration in wireless system is usually fixed, therefore we need to modify the proposed scheme to be effectively used in wireless systems. For this aim, we assume that the time slot duration is fixed and equals $d_p$. This will be equivalent to the original random NOMA scheme with dynamic time slot duration, when a delay constraint is imposed to the system. In this case, the devices need to be decoded at the base station with a delay of at most $d_p$, which can be also translated to a system with a fixed time slot duration of length $d_p$, where the devices are allowed to transmit in only one time slot. In this section, we find the stability condition for the proposed scheme under a delay constraint and characterize the maximum throughput which can be supported by the proposed scheme under the delay constraint.

Let $n_1$ denote the number of active devices at the first time slot. An active device can then deliver its message with delay $T(n_1)$ with probability $1-\mathrm{P_c}(n_1)$; otherwise it will reattempt the transmission in the next time slot. More specifically, the device's message can be delivered at the BS at the $j^{th}$ time slot with the probability given below:
\begin{align}
\mathbb{P}[I=j|n_1]&=\sum_{(n)_{2}^{j}}\left(1-\mathrm{P_c}(n_{j})\right)\prod_{i=1}^{j-1}\mathrm{P_c}(n_{i})\mathbb{P}[n_{i+1}|n_{i}].
\label{eq:delayindex}
\end{align}
Let us assume that the BS can change the number of seeds such that the collision probability is always less than $p_c$ regardless of the number of active devices. Then (\ref{eq:delayindex}) is reduced to:
\begin{align}
\mathbb{P}[I=j]\approx(1-p_c)p_c^{j-1},
\label{eq:simplifiedtimeindex}
\end{align}
which is a decreasing function of $j$. The delay can then be characterized as follows:
\begin{align}
\nonumber \mathbb{P}[d|\lambda, n_1]&=\sum_{j}\mathbb{P}[d|I=j]\mathbb{P}[I=j]\\
&=\sum_{j}\mathbb{P}[I=j]\sum_{(n)_{2}^{j}}\mathbb{P}[d|n1,\cdots,n_j]\prod_{i=1}^{j-1}\mathbb{P}[n_{i+1}|n_i]\\
&=\sum_{j}\mathbb{P}[I=j]\sum_{(n)_{2}^{j}}\bigotimes_{i=1}^{j}\mathbb{P}[d_i|n_i]\prod_{i=1}^{j-1}\mathbb{P}[n_{i+1}|n_i],
\label{eq:delay}
\end{align}
where $\bigotimes$ is the convolution operator,
\[\mathbb{P}[d_i|n_i]=\left\{\begin{array}{ll}
q(c_i,n_i),~~&d_i=\frac{L}{W_s\log_2\left(1+\frac{\mu}{1+(c_i-1)\mu}\right)},\\
0,~~&\text{otherwise,}
\end{array}\right.\]
and $\mathbb{P}[n_i|n_{i-1}]$ is given in (\ref{exactPoissBinom}). By using (\ref{eq:simplifiedtimeindex}) and considering only the first few terms of (\ref{eq:delay}), i.e., $j=1,2$, (\ref{eq:delay}) can be simplified as follows:
\begin{align}
\nonumber\mathbb{P}[d|n_1]&\approx(1-p_c)\mathbb{P}[d|n_1]\\
&+p_c(1-p_c)\sum_{n_2}\frac{\mathbb{P}[d|n_1]\otimes \mathbb{P}[d|n_2]}{\sqrt{2\pi\sigma^2_2}}~e^{-\frac{(n_2-\mu_2)^2}{\sigma^2_2}},
\label{eq:approxdelay}
\end{align}
where $\mu_2=\lambda \overline{T}(n_1)+np_c$ and $\sigma^2_2=\lambda \overline{T}(n_1)+n_1p_c(1-p_c)$.

The weak stability condition is defined as follows:
\begin{align}
\mathbb{E}[d|\lambda,n_1]\le d_p.
\end{align}
As shown in (\ref{eq:weaknodelay}), the maximum arrival rate under the weak stability condition is given by $n(1-p_c)/\overline{T}(n)$. Therefore, the maximum initial backlog under the weak stability condition to satisfy the delay requirement $d_p$ is given by:
\begin{align}
n^{(\mathrm{weak,delay})}_{\max}(d_p)=\max_{n}\left\{n\left|\mathbb{E}\left[d\left|n,\frac{n(1-p_c)}{\overline{T}(n)}\right.\right]\le d_p\right.\right\}
\end{align}
and by using (\ref{eq:weaknodelay}), the maximum packet arrival rate under the weak stability condition is given by:
\begin{align}
\lambda^{(\mathrm{weak,delay})}_{\max}(d_p)=\frac{n^{(\mathrm{weak})}_{\max}(d_p)(1-p_c)}{\overline{T}(n^{(\mathrm{weak,delay})}_{\max}(d_p))}.
\end{align}

Similarly, the strong stability condition can be found as follows:
\begin{align}
1-\int_{0}^{d_p}\mathbb{P}[d=\tau|n,\lambda]d\tau<\epsilon,
\end{align}
the maximum initial backlog under the strong stability condition is given by
\begin{align}
\nonumber &n^{(\mathrm{strong,delay})}_{\max}(d_p)\\
&=\max_{n}\left\{n\left|1-\int_{0}^{d_p}\mathbb{P}\left[d=\tau\left|n,\lambda^{(strong)}_{\max}(n)\right.\right]d\tau<\epsilon\right.\right\},
\end{align}
where $\lambda^{(strong)}_{\max}(n)$ is given in (\ref{eq:strongcloseform}). The maximum packet arrival rate under the strong stability condition is then given by:
\begin{align}
\lambda^{(strong,delay)}_{\max}(d_p)=\lambda^{(strong)}_{\max}(n^{(\mathrm{strong,delay})}_{\max}(d_p)).
\end{align}

\section{Numerical Results}
\begin{figure}
  \centering
  \includegraphics[width=0.5\columnwidth]{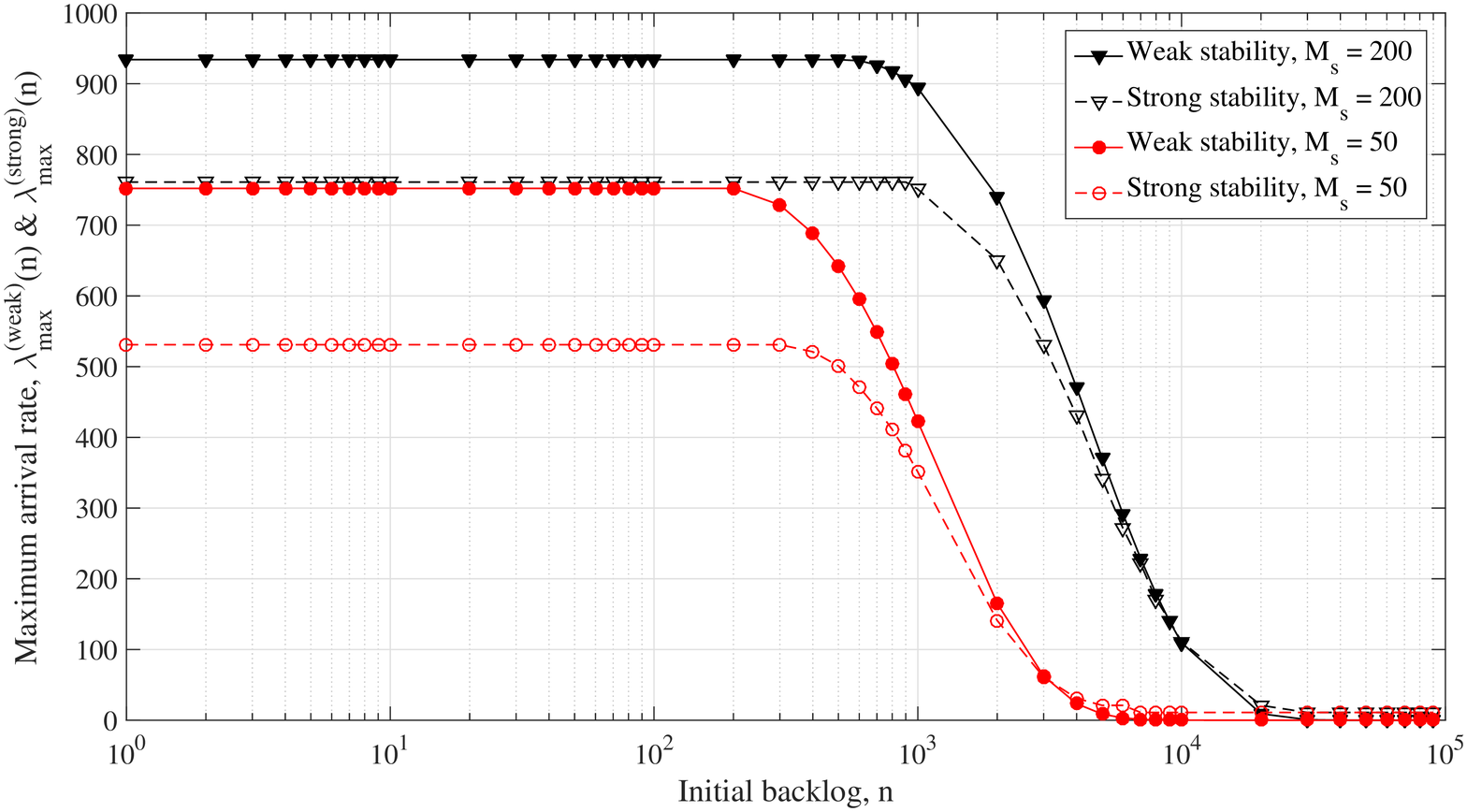}
  \vskip-1.5ex
  \caption{The maximum arrival rate versus the initial backlog obtained from the weak and strong stability conditions for different $M_s$, when $N_s=20$, $W=1$ MHz, $L=1000$, and the threshold probability for the strong stability condition is $\epsilon=0.01$.}
  \label{fig:stabMs}
  \end{figure}
\begin{figure}
  \centering
  \includegraphics[width=0.5\columnwidth]{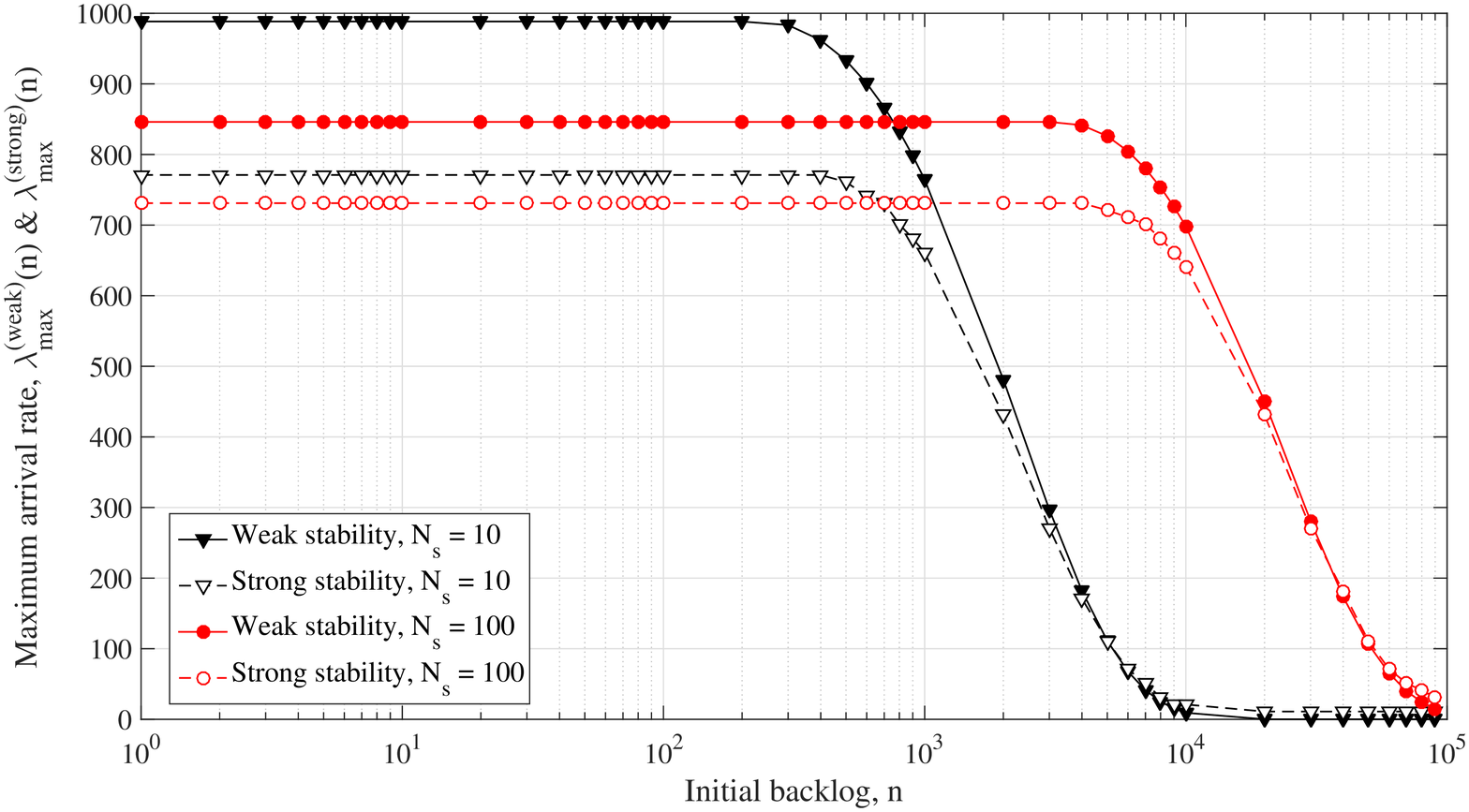}
  \vskip-1.5ex
  \caption{The maximum arrival rate versus the initial backlog obtained from the weak and strong stability condition for different $N_s$, when $M_s=200$, $W=1$ MHz, $L=1000$, and the threshold probability for the strong stability condition is $\epsilon=0.01$.}
  \label{fig:stabNs}
\end{figure}
Fig. \ref{fig:stabMs} shows the maximum arrival rate versus the initial backlog $n$, for different numbers of seeds, $M_s$, when the total available bandwidth is $W=1$ MHz, the number of sub-bands is $N_s=20$, and each device attempts to deliver a message of length $L=1000$ bits to the BS. As can be seen, the system with higher $M_s$ can support more devices as the collision probability decrease with $M_s$; so fewer devices will reattempt their transmissions in the following time slot.

Fig. \ref{fig:stabNs} shows the stability regions for different number of sub-bands when the number of seeds is $M_s=200$. With increasing number of sub-bands, the collision probability decreases but on the other hand the bandwidth of each sub-band will also decrease. This results in longer time slots as the transmission of the devices takes longer due to smaller bandwidth and lower achievable rate over each sub-band. As can be seen in Fig. \ref{fig:stabNs}, with increasing $N_s$, the maximum arrival rate decreases but the system can support a larger initial backlog.

\begin{figure}[t]
  \centering
  \includegraphics[width=0.5\columnwidth]{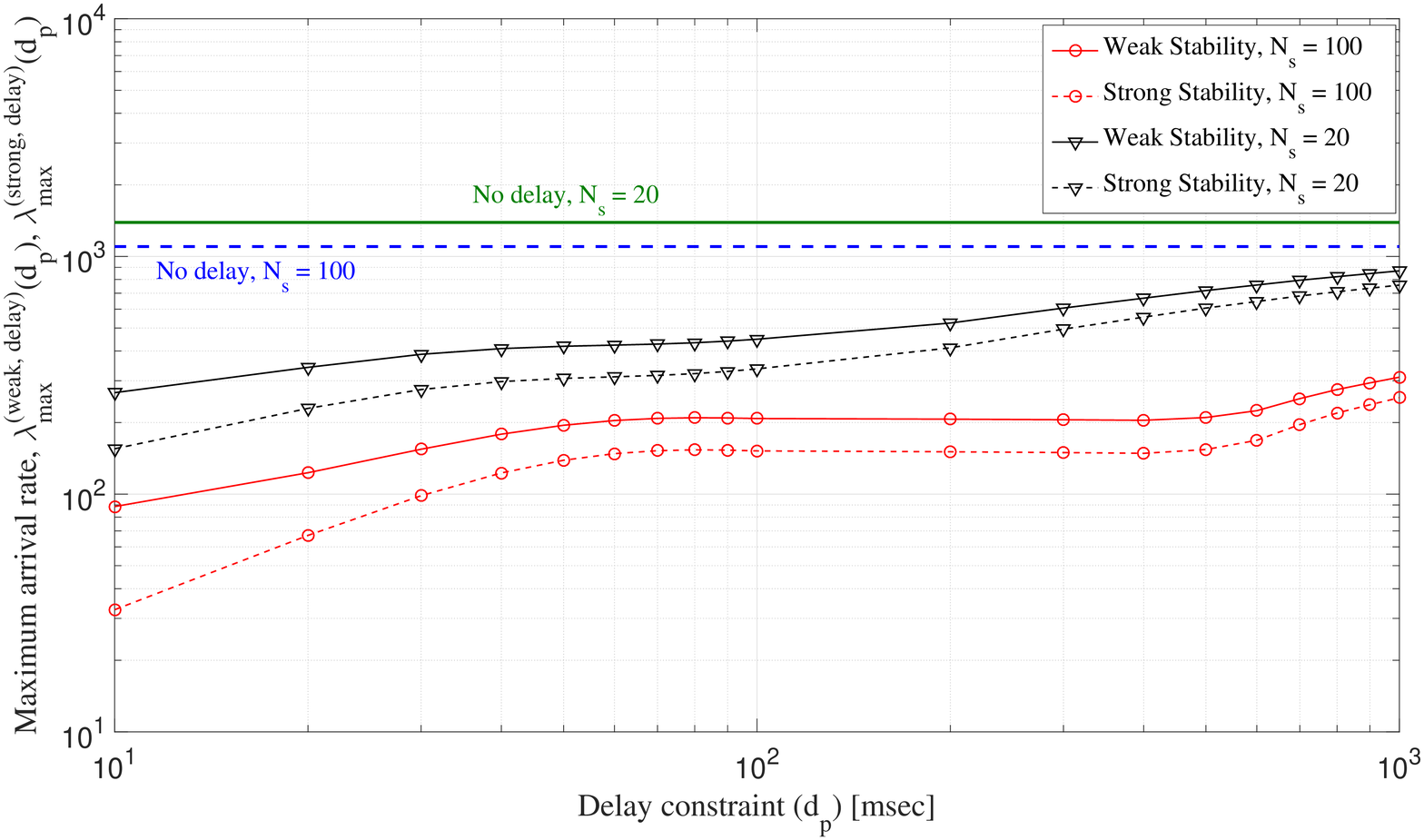}
  \vskip-1.5ex
  \caption{The maximum packet arrival rate versus the delay constraint under weak and strong stability conditions, when $W=1$ MHz, $L=1000$, and collision probability is set to $p_c=0.01$. The threshold probability for the strong stability condition is $\epsilon=0.01$.}
  \label{fig:delaystability}
\end{figure}

Fig. \ref{fig:delaystability} shows the maximum arrival rate versus the delay constraint under weak and strong stability conditions. As can be seen, in delay sensitive conditions, i.e., short delay, the number of devices which can be supported by the proposed NOMA strategy is small, and by increasing the tolerable delay, the supported arrival rate increases. For comparison, we have also shown the maximum arrival rate without a delay constraint in Fig. \ref{fig:delaystability}. As can be seen by relaxing the delay constraint the maximum supported arrival rate gets closer to the maximum supported arrival rate without the delay constraint.  It is important to note that when the number of subbands is large, e.g., $N_s=100$, the available bandwidth for each subband is very small, and only a very small number of devices can be supported to satisfy the delay constraint, which explains the large gap between the the maximum supported arrival rate of a system without and with delay constraints.

\section{System Optimization}
\subsection{Number of sub-bands}
The number of sub-bands and available seeds will determine the overall system performance as the collision probability and the maximum achievable rate for the proposed random NOMA strategy will be mainly determined by these two parameters. The base station then needs to find the optimal values for these parameters to maximize the system throughput or satisfy the QoS requirements of the devices. It is clear that the collision probability decreases as $M_s$, the number of seeds, increases. One could adaptively change the number of seeds according to the incoming traffic at the BS to fix the collision probability. However, it is also clear that increasing the number of seeds adds extra complexity at the BS as the BS should consider a larger number of seeds while performing SIC.

We first consider the optimization of the supported arrival rate when there is no delay constraint. The maximum supported arrival rate according to the strong stability condition without delay constraint, i.e., (\ref{eq:strongcloseform}), for a given $\epsilon$, $M_s$, $W$ and $L$, is given by:
\begin{align}
\max_{\{N_s,n\}}\frac{1+2n\ell_{\epsilon}(1-\mathrm{P_c}(n))-\sqrt{1+4n\ell_{\epsilon}(1-\mathrm{P_c}(n))^2}}{2\ell_{\epsilon}\overline{T}(n)},
\end{align}
where $\overline{T}(n)$, $\mathrm{P_c}$, and $\ell_{\epsilon}$ are respectively given by (\ref{eq:AndTimeDuration}), (\ref{eq:collisionProb}), and (\ref{eq:ellepsilon}).

\begin{figure}
  \centering
  \includegraphics[width=0.5\columnwidth]{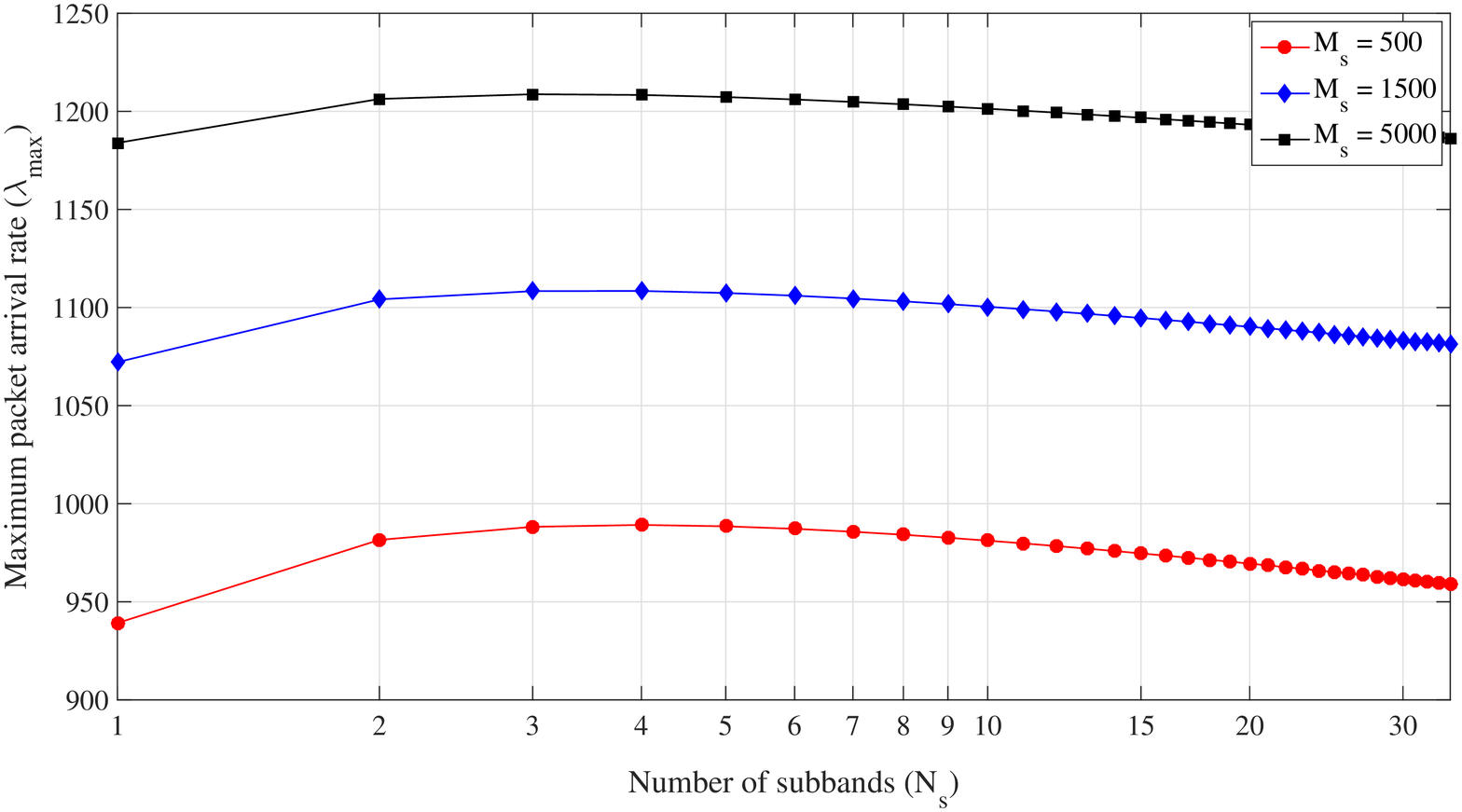}
    \vskip-1.5ex
  \caption{The maximum packet arrival rate under the strong stability condition without delay constraint versus the number of sub-bands, $N_s$, when $W=1$ MHz, $L=1000$, and the threshold probability for the strong stability condition is $\epsilon=0.01$.}
  \label{fig:maxNs}
\end{figure}
\begin{figure}
  \centering
  \includegraphics[width=0.5\columnwidth]{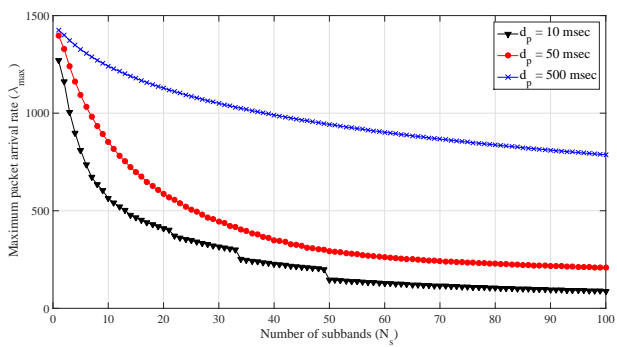}
    \vskip-1.5ex
  \caption{The maximum packet arrival rate under the strong stability condition with delay constraint versus the number of sub-bands, $N_s$, when $W=1$ MHz, $L=1000$, and the threshold probability for the strong stability condition is $\epsilon=0.01$.}
  \label{fig:maxNsdelay}
\end{figure}
Fig. \ref{fig:maxNs} shows the maximum packet arrival rate versus the number of sub-bands without a delay constraint. As can be seen in this figure, the maximum packet arrival rate is achieved when the number of sub-bands is either 3 or 4 for different numbers of seeds. This means that to support a large number of devices using the proposed random NOMA strategy, the available bandwidth does not need to be divided into too many sub-bands, only a few sub-bands is sufficient. This is because when the number of sub-bands increases, the available bandwidth for each sub-band decreases, which also decreases the maximum achievable rate over each sub-band; therefore, fewer packets will be delivered over each sub-band.

A similar optimization problem can be defined to find the optimal number of sub-bands to maximize the supported arrival rate for a given delay constraint. Fig. \ref{fig:maxNsdelay} shows the maximum packet arrival rate versus the number of sub-bands for different delay constraints $d_p$. As can be seen, for a given $d_p$, the maximum packet arrival rate can be supported when the whole bandwidth is used as only one sub-band. In other words, dividing the bandwidth into several sub-bands degrades the performance of the proposed random NOMA in terms of the packet arrival rate which can be supported at the BS within a given delay requirement. Therefore, to satisfy the QoS requirements of a large number of devices using the proposed random NOMA strategy, the devices should use the whole bandwidth and the BS should control the collision probability by choosing a larger seed pool.
\subsection{MTC Device Fairness}
\begin{figure}[t]
  \centering
  \includegraphics[width=0.5\columnwidth]{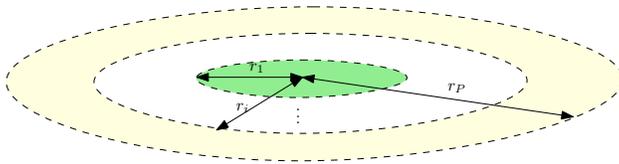}
  \vskip-1.5ex
  \caption{Dividing a cell into $N_s$ partitions.}
  \label{fig:cellpartition}
\end{figure}

In the proposed random NOMA strategy, we have assumed that the signals received from all the devices have the same power at the BS. This way, the devices which are far from the BS should transmit with higher power to maintain the same received power at the BS. In other words, the devices which are far from the BS should spend more energy to achieve the same throughput performance as the devices close to the BS. One approach to solve this problem is to allocate bandwidth to the devices according to their distances to the BS, so they can achieve the same throughput performance with the same energy consumption. For this aim as shown in Fig. \ref{fig:cellpartition}, we divide the cell into $N_s$ partitions, such that the area covered in each partition is the same. This way the average number of devices in each partition is the same due to the fact that the devices are randomly distributed in the cell. Let $r_i$ denote the radius of the outer edge of the $i^{th}$ partition for $i=1,\cdots,N_s$, where $r_{N_s}=R_o$. Then, it is easy to show that $r_i$ is given by:
\begin{align}
r_i=\sqrt{\frac{i}{N_s}}R_o.
\label{eq:radius}
\end{align}

In particular, each device estimate its location based on its channel condition which is estimated using  a pilot signal regularly transmitted by the base station. The BS broadcasts the information about the partitions to the devices. Each partition is characterized according to the maximum and minimum average received power at the base station. And these values are broadcasted to the devices, so each device knows the cell partition it belongs to by comparing its received power with the threshold powers.

Unlike the original random NOMA presented in Section III, where the devices randomly choose among $N_s$ available sub-bands of the same bandwidth and their received power at the BS is the same, here we assume that the total bandwidth is divided into $N_s$ sub-bands, where the devices in the $i^{th}$ cell partition transmit in the $i^{th}$ sub-band with bandwidth $W_i$ such that their received SNR at the BS is $\mu_i$. The non-uniform allocation of the bandwidth to the sub-bands allows for the derivation of a fair multiple access strategy in terms of the energy consumption, which is explained in the following.

To have a fair system, we need to guarantee the same average throughput and energy consumption for all the devices regardless of their distances to the BS. To achieve the same average throughput over all the sub-bands and accordingly all the cell partitions, the average duration of the sub-bands should be the same.

\begin{lemma}
Under the fairness constraint, where the average time slot duration over all the subbands are the same and the average energy consumption for all the devices are the same, the bandwidth for each subband, which also corresponds to a cell partition, should be allocated as follows,
\begin{align}
W_i=\frac{i^{\frac{\alpha+2}{2}}-{(i-1)}^{\frac{\alpha+2}{2}}}{N_s^{\frac{\alpha+2}{2}}}W.
\end{align}
where $W_i$ is the bandwidth of the $i^{th}$ subband and $W$ is the total available bandwidth.
\end{lemma}

\begin{IEEEproof}
See Appendix D.
\end{IEEEproof}

\begin{figure}[t]
  \centering
  \includegraphics[width=0.5\columnwidth]{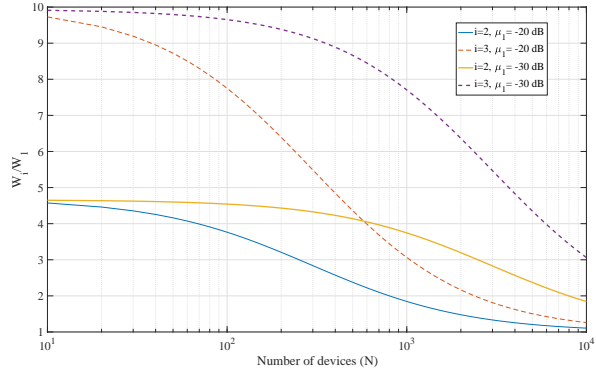}
  \vskip-1.5ex
  \caption{Fair bandwidth allocation for energy efficient massive NOMA transmission. The total bandwidth is $W=1$ MHz, path loss exponent is $\alpha=3$, and $N_s=3$. $\mu_i$ can be calculated for different $i$ using (\ref{eq:mucalc}).}
  \label{fig:OptBandwidth}
\end{figure}

This shows that, to have the same energy consumption for all the devices across the cell, more bandwidth should be allocated to those devices which are far from the BS. Fig. \ref{fig:OptBandwidth} shows the bandwidth allocation versus the number of devices when the total bandwidth is 1 MHz and the number of sub-bands (or equivalently the number of cell partitions) is 3. As can be seen, with increasing the number of devices, the bandwidth will be allocated more evenly between the sub-bands, which is because $n$ is the dominant term in (\ref{eq:bandw1}) when $n$ is very large. On the other hand, when $n$ is relatively small, more bandwidth is allocated to the devices which are located far from the BS. This shows that the BS needs to have a proper load estimation strategy to allocate the bandwidth between the sub-bands so as to obtain fairness in the energy consumption and throughput.

\section{Practical Considerations of Massive NOMA for M2M Communications}
Although NOMA can improve spectrum efficiency and system capacity, there are many practical challenges for this technology to be potentially used in real wireless systems for M2M communications. A summary of most important challenges of NOMA has been presented in \cite{mag_Mahyar}. Here we emphasize three main challenges of massive NOMA and propose some solutions to effectively solve them and take an step towards developing a more practical massive IoT system using the NOMA strategy.

\subsection{System Overload}
when a large number of devices are transmitting simultaneously. Once the base station detects that the system is overloaded, i.e., the number of arrivals is larger than the maximum supportable number of devices, it can either increase the number of seeds to minimize the collision probability or let some of the users not transmit in a particular time slot. This way the base station can effectively control the traffic and distribute the traffic over time. In fact back-off and access barring strategies which are commonly used in cellular systems can be combined with the proposed scheme to handle the overload traffic and effectively delay the traffic. In fact, this has no effect on a system without delay constraint, however, it reduces the throughput when a strict delay constraint is imposed to the system.
\subsection{Optimal Power Allocation and Throughput}
As we mentioned earlier, the duration of a sub-band is determined by the rate achieved by the device with the lowest SINR, as we assumed that devices' message are received with the same SNR over each sub-band. The minimum rate achieved by the devices in a sub-band containing $c$ devices is given by:
\begin{align}
R_{\min}=\log_2\left(1+\frac{\mu_0}{1+(c-1)\mu_0}\right).
\end{align}
and the effective rate of the device which is decoded in the $j^{th}$ stage of the SIC process is given by:
\begin{align}
R_{j}=\log_2\left(1+\frac{\mu_0}{1+(c-j)\mu_0}\right).
\end{align}
However, as the devices do not know the traffic load and randomly transmit over the sub-bands in a rateless manner, the effective rate cannot be achieved by the devices as all devices must transmit at rate $R_{\min}$. This means that the optimal throughput cannot be achieved, which is mainly due to the fact that the devices are unknown to the BS, so it cannot optimally determine the power and rates. The SIC order is not relevant here, as all the devices perform over the same data rate.

To achieve the full potential of NOMA, the devices' messages need to be received with different powers, where the power levels are determined by the BS to optimize the throughput. However, this is only practical when the BS can identify the devices before the data transmission so it can optimally determine their received power. That is, if the devices can exchange more information to the BS before their data transmission or ideally be identified at the BS, the BS can determine optimal transmission strategy in terms of power and rate and broadcast this information to the devices. This is, however, impractical for massive IoT applications where the message size is usually small, so the overhead must be kept as small as possible. On the other hand, it would be impractical for the BS to identify and perform channel estimation to a large number of devices in each time instant. This would incur huge delay in the system which is not acceptable for most massive IoT applications. The solution would be to minimize the control overhead by removing the device identification phase (as in the proposed scheme), and improve the system throughput by optimizing the bandwidth allocation as discussed in Section Vi.
\subsection{Delay Imposed from the SIC Process}
\begin{figure}[t]
  \centering
  \includegraphics[width=0.4\columnwidth]{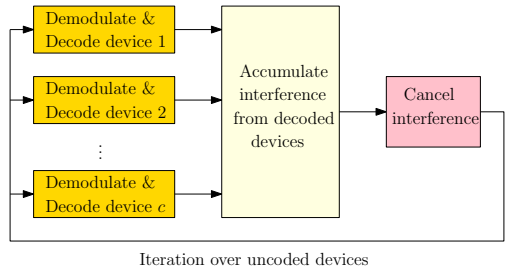}
  \vskip-1.5ex
  \caption{Parallel interference cancellation for the proposed random NOMA strategy.}
  \label{fig:parIC}
\end{figure}

In the proposed random NOMA strategy, we consider the successive interference cancellation at the BS, that is the BS starts the decoding of the device with the highest SINR and then removes its interference from the received signal and continues the decoding of the remaining devices. However, this imposed some delay into the system as a device should wait some time for the previous devices to be decoded by the BS before being decoded in the SIC process depending on the decoding order chosen by the BS. As all the devices are assumed to transmit with the same data rate over the same subband, parallel decoding can be performed to decode the messages. Also, to address the delay issue in SIC, we can consider iterative parallel interference cancellation \cite{patel1994performance,divsalar1994improved} shown in Fig. \ref{fig:parIC}. This scheme is more attractive from an implementation perspective as multiple devices are decoded and cancelled from the received signal simultaneously.

We can also assume that delay sensitive devices use specific random seeds, and accordingly group the devices based on their delay requirements; this could also be beneficial from an energy consumption point of view \cite{Condoluci20163,7389317}. By applying this concept to our proposal, then Parallel IC is performed on high priority groups (correspond to low latency devices) to low priority groups (correspond to delay tolerant devices) successively \cite{LinearGroup}. The readers are referred to \cite{verdu1998multiuser,patel1994performance,divsalar1994improved,hui1998successive,wang1999iterative} for further details.
\subsection{A brief comparison between Narrow-band IoT and Masive NOMA}
As part of 3GPP Release 13, narrow-band IoT (NB-IoT) \cite{nbiot3gpp,3gpp_36211,3gpp_36212,3gpp_36300} has been standardized for low end massive IoT, that is the devices require relatively low data rate ($\sim$250 kbps in downlink direction, $\sim$20 kbps in uplink with the possibility to aggregate multiple tones to reach the same speed as in downlink) with relaxed delay requirements (in the order of 10 seconds). The required bandwidth for NB-IoT is 180 KHz for both uplink and downlink.

NB-IoT promises to improve the cellular systems for massive IoT by supporting of massive number of low throughput devices, where up to 50000 devices can be supported per cell, for the arrival traffic of about 6 packets per second. Despite this, NB-IoT is still based on a two-step procedure (i.e,. random access followed by data transmission) which is only appropriate for low packet arrival rates as it limits the overall channel capacity. Moreover, as the delay assumption was relaxed in NB-IoT, it does not provide a solution for devices with strict delay requirements.\footnote{For delay-constrained applications, 3GPP proposed a further Long Term Evolution (LTE) enhancement for machine-type communications, i.e., eMTC.}

As a simple comparison with NB-IoT, we consider the total system bandwidth of $W=180$ kHz and 12 subbands each of 15 kHz bandwidth. Our simulations show that NOMA can support arrival rates up to 100 packets per second under the strict delay requirement of 100 msec and arrival rates of up to 180 packets per second for the delay requirement of 1 sec. This is much more than what can be supported by NB-IoT which supports around 18 packets per second\footnote{It is worth mentioning that the target capacity for NB-IoT is about 55000 devices per cell sector, which corresponds (by considering devices transmitting one packet per hour) to a target traffic of $\sim$15 packets per second.}, with a uplink latency up to $\sim$2.8 s for devices with very poor channel coverage \cite{nbiot3gpp}. This shows that NOMA can be used for delay sensitive applications and support a larger number of devices compared to NB-IoT.

\section{Conclusions}
We considered a random non-orthogonal multiple access strategy for massive IoT, where multiple devices are allowed to transmit over the same sub-band and the base station performs successive interference cancellation to decode each device's message. We derived system stability conditions, where the maximum packet arrival rate was found with and without quality of service guarantee. We then found the optimized system parameters, including the number of sub-bands under these scenarios; optimizing the throughput alone, including a delay constraint, and ensuring user fairness in both throughput and energy consumption. We found that the optimal strategy differed for each of these conditions. More specifically, we found that without any delay constraint the whole bandwidth must be divided into only a few (i.e., 3 or 4) sub-bands to maximize the packet arrival rate which can be supported by the base station. On the other hand, when a delay constraint is imposed on the system, the whole bandwidth must be used as only one sub-band to support a large packet arrival rate and satisfy the delay requirement.

\appendices

\section{Proof of Lemma 1}
Let $k_i$ denote the number of devices which have selected the $i^{th}$ sub-band, where we have dropped the time index for the ease of notation. Let $q(c,n)$ denote the probability that the maximum number of devices over all the sub-bands is $c$ when the number of active devices is $n$. It is then easy to shows that $q(c,n)$ is given by:
\begin{align}
q(c,n)=\frac{\left|\left\{(k)_{1}^{N_s}\left|\displaystyle\sum_{j=1}^{N_s}k_i=n\right., \displaystyle\max_j k_j=c\right\}\right|}{N_s^n}.
\label{eq:exactexpqc}
\end{align}

For sufficiently large $n$ and $N_s$, we can approximate the number of devices in each sub-band by a binomial distribution. This is due to the fact that each device randomly and independently selects among $N_s$ available sub-bands with equal probability. More specifically, the probability mass function ($\mathrm{p.m.f.}$) of $k_i$ for $i=1,\cdots, N_s$ is given by:
\begin{align}
\mathbb{P}[k_i|n]=\dbinom{n}{k_i}\left(\frac{1}{N_s}\right)^{k_i}\left(1-\frac{1}{N_s}\right)^{n-k_i},
\end{align}
which can be further approximated by the normal distribution as follows \cite{papoulis2002probability}:
\begin{align}
\mathbb{P}[k_i|n]\approx \frac{1}{\sigma_s}\varphi\left(\frac{k_i-\frac{n}{N_s}}{\sigma_s}\right),
\end{align}
where $\varphi(x)=\frac{1}{\sqrt{2\pi}}e^{-\frac{x^2}{2}}$ and $\sigma_s=\sqrt{\frac{n}{N_s}(1-\frac{1}{N_s})}$. We aim at finding the $\mathrm{p.m.f.}$ of the maximum number of devices over all the available sub-bands. We first derive its cumulative mass function ($\mathrm{c.m.f.}$) as follows:
\begin{align}
\nonumber \mathbb{P}&\left[\left.\max_i\{k_i\}\le \ell\right|n\right]=\mathbb{P}\left[\left.k_1\le\ell, k_2\le\ell,\cdots, k_{N_s}\le\ell\right|n\right]\\
\nonumber&=\prod_{i=1}^{N_s}\mathbb{P}\left[k_i\le\ell|n\right]=\mathbb{P}[k_1\le\ell|n]^{N_s}\approx \left[1-\Phi\left(\frac{\ell-\frac{n}{N_s}}{\sigma_s}\right)\right]^{N_s},
\end{align}
where $\Phi(x)=\int_{x}^{\infty}\varphi(t)dt$ is the cumulative distribution function ($\mathrm{c.d.f.}$) of the standard normal distribution. The $\mathrm{p.m.f.}$ of the maximum number of devices over all the sub-bands can then be derived as follows:
\begin{align}
q(c,n)\approx\frac{N_s\varphi\left(\frac{c-\frac{n}{N_s}}{\sigma_s}\right)\left[1-\Phi\left(\frac{c-\frac{n}{N_s}}{\sigma_s}\right)\right]^{N_s-1}}{\sigma_s}
\label{eq:finalAppqc}
\end{align}

\section{Proof of Lemma 2}
Let $n_i$ denote the number of active devices in the $i^{th}$ time slot, where $i\ge1$. The probability that $n_{i+1}$ devices are attempting to deliver their messages to the BS in the ${(i+1)^{th}}$ time slot is given by:
\begin{align}
\mathbb{P}[n_{i+1}|n_i]=\sum_{j=0}^{\min\{n_{i+1},n_i\}}e^{-\lambda \overline{T}(n_i)}\frac{(\lambda \overline{T}(n_i))^{n_{i+1}-j}\dbinom{n_i}{j}\mathrm{P_c}(n_i)^j}{(n_{i+1}-j)!(1-\mathrm{P_c}(n_i))^{j-n_{i}}}.,
\label{exactPoissBinom}
\end{align}
In fact, $j$ out of $n_{i+1}$ devices might be those packets which have collided in the $i^{th}$ time slot, while the remaining $(n_{i+1}-j)$ packets are newly generated packets. The number of collided packets is a random variable which follows a binomial distribution with success probability $\mathrm{P_c}(n_i)$, as each device is independently in collision with probability $\mathrm{P_c}(n_i)$, which is true when the number of devices is sufficiently large. It can be further approximated by a normal distribution (see (4-35) in \cite{papoulis2002probability}) with mean $n_i\mathrm{P_c}(n_i)$ and variance $n_i(1-\mathrm{P_c}(n_i))$ \cite[equation 4-95]{papoulis2002probability}. The number of newly generated packets is also a random variable which is assumed to follow a Poisson distribution with mean $\lambda$ $\mathrm{packets/sec}$, which can be also approximated by a normal distribution with mean and variance $\lambda \overline{T}(n_i)$, when $\lambda \overline{T}(n_i)$ is sufficiently large \cite[equation 4-107]{papoulis2002probability}. These random variables are mutually independent, therefore, the probability that $n_{i+1}$ devices are transmitting in the $(i+1)^{th}$ time slot can be calculated by multiplying the probability of $j$ collided devices and $n_{i+1}-j$ newly generated packets and taking the summation over $j$.

Using normal approximations for the number of collided devices and newly generated packets, (\ref{exactPoissBinom}) can be simplified as follows:
\begin{align}
\mathbb{P}[n_{i+1}|n_i]\approx \frac{\exp\left(-\frac{(n_{i+1}-\mu_i)^2}{\sigma^2_i}\right)}{\sqrt{2\pi\sigma_i^2}},
\label{AppPoissBinom}
\end{align}
where $\mu_i=\lambda \overline{T}(n_i)+n_i\mathrm{P_c}(n_i)$ and $\sigma^2_i=\lambda \overline{T}(n_i)+n\mathrm{P_c}(n_i)(1-\mathrm{P_c}(n_i))$. The average number of devices in the $(i+1)^{th}$ time slot is then given by:
\begin{align}
\mathbb{E}[n_{i+1}|n_i]=\lambda \overline{T}(n_i)+ n\mathrm{P_c}(n_i).
\end{align}

Under the weak stability condition (\ref{weakstabcond}), for a system with backlog $n$ we have
\begin{align}
\lambda \overline{T}(n)+ n\mathrm{P_c}(n) \le n,
\end{align}
and the maximum arrival rate can be easily characterized by (\ref{eq:weaknodelay}). The maximum arrival rate which can be supported by the system in then found using (\ref{weakfinalcond}).

\section{Proof of Lemma 3}
By using (\ref{exactPoissBinom}), the strong stability condition can be derived as follows:
\begin{align}
1-\sum_{n'=0}^{n}\sum_{i=0}^{n'}e^{-\lambda \overline{T}(n)}\frac{(\lambda \overline{T}(n))^{n'-i}\dbinom{n}{i}\mathrm{P_c}^i}{(n'-i)!(1-\mathrm{P_c})^{i-n}} \le \epsilon,
\end{align}
which can be also written as follows by using (\ref{AppPoissBinom}):
\begin{align}
\Phi(\frac{n-\mu}{\sigma})\le \epsilon,
\label{eq:maxstrong}
\end{align}
where $\mu=\lambda \overline{T}(n)+n\mathrm{P_c}(n)$ and $\sigma^2=\lambda \overline{T}(n)+n\mathrm{P_c}(n)(1-\mathrm{P_c}(n))$. One could easily find the maximum arrival rate using (\ref{eq:maxstrong}) as follows:
\begin{align}
\Phi^{-1}(\epsilon)\le \frac{n-\lambda \overline{T}(n)-n\mathrm{P_c}(n)}{\sqrt{\lambda \overline{T}(n)+n\mathrm{P_c}(n)(1-\mathrm{P_c}(n))}},
\end{align}
and by solving this inequality with respect to $\lambda$ we can easily derive (\ref{eq:strongcloseform}). The maximum arrival rate can then be easily found by maximizing over $n$ as in (\ref{eq:strongwodelay}).

\section{Proof of Lemma 4}
As the devices are randomly distributed in the cell, $j$ out of $n$ active devices belong to the $i^{th}$ cell partition with the probability given below:
\begin{align}
\mathbb{P}[n_i=j|n]=\dbinom{n}{j}\left(\frac{1}{P}\right)^{j}\left(1-\frac{1}{P}\right)^{n-j},
\end{align}
and the time required for these devices to deliver their messages at the BS in the $i^{th}$ sub-band is given by (\ref{eq:timeduration}):
\begin{align}
t_{i}(j)=\frac{L}{W_i\log_2\left(1+\frac{\mu_i}{1+(j-1)\mu_i}\right)}.
\end{align}

The average duration of the $i^{th}$ sub-band can then be calculated as follows:
\begin{align}
\overline{T}_i(n)=\sum_{j=0}^{n}\frac{L\dbinom{n}{j}\left(\frac{1}{N_s}\right)^{j}\left(1-\frac{1}{P}\right)^{n-j}}{W_i\log_2\left(1+\frac{\mu_i}{1+(j-1)\mu_i}\right)}.
\end{align}
This can be simplified for $\mu_i$ being sufficiently small by using the first term of the Maclaurin series $\ln(1+x)=x+\mathcal{O}(x^2)$ assuming that $x$ is very small:
\begin{align}
\nonumber\overline{T}_i(n)&=\sum_{j=0}^{n}\frac{L\ln(2)\dbinom{n}{j}\left(\frac{1}{N_s}\right)^{j}\left(1-\frac{1}{N_s}\right)^{n-j}}{W_i\frac{\mu_i}{1+(j-1)\mu_i}}\\
\nonumber &=\frac{L\ln(2)}{W_i\mu_i}\sum_{j=0}^{n}\dbinom{n}{j}\left(\frac{1}{N_s}\right)^{j}\left(1-\frac{1}{N_s}\right)^{n-j}\left(1+\mu_i(j-1)\right)\\
&\overset{(a)}=\frac{L\ln(2)}{W_i\mu_i}\left(1+\mu_i\left(\frac{n}{N_s}-1\right)\right),
\label{eq:appavgt}
\end{align}
where step $(a)$ follows from the fact that $\sum_{j=0}^{n}\dbinom{n}{j}(1/N_s)^j(1-1/N_s)^{n-j}=1$ and the mean value of a Binomial distribution with success probability $1/N_s$ is $\sum_{j=0}^{n}j\dbinom{n}{j}(1/N_s)^j(1-1/N_s)^{n-j}=n/N_s$. In order to have the same average time duration for all the sub-bands, we need to satisfy $\overline{T}_i(n)=\overline{T}_1(n)$ for $i=1,\cdots, N_s$, which can be rewritten as follows using (\ref{eq:appavgt}):
\begin{align}
\frac{W_i}{W_1}=\frac{\mu_i^{-1}+\left(\frac{n}{N_s}-1\right)}{\mu_1^{-1}+\left(\frac{n}{N_s}-1\right)}\approx \frac{\mu_1}{\mu_i},
\label{eq:bandw1}
\end{align}
where the approximation follows from the assumption that the $\mu_i$'s are very small.

To maintain the same average energy consumption for all the devices, the average transmit power for all the devices should be the same given that we have required that the average duration of the sub-bands are the same. By using (\ref{eq:transmitpower}), the average transmit power to achieve SNR $\mu_i$ over the $i^{th}$ sub-band is given by:
\begin{align}
\nonumber\overline{P}_{\mathrm{t},i}&=P_{\max}\frac{\mu_i}{\mu_{\mathrm{ref}}\chi h}\int_{r_{i-1}}^{r_i}\left(\frac{r}{R_\mathrm{o}}\right)^{\alpha}\frac{2r}{R^2_o}dr\\
&=P_{\max}\frac{\mu_i}{\mu_{\mathrm{ref}}\chi h}\frac{r_i^{\alpha+2}-r_{i-1}^{\alpha+2}}{(2+\alpha)R_o^{\alpha+2}}.
\label{eq:transmitpower1}
\end{align}
The average energy consumption of the devices in the $i^{th}$ cell partition, which are transmitting in the $i^{th}$ sub-band, is given by $\overline{E}_i=\overline{T}_i(n)\overline{P}_{\mathrm{t},i}$. As we assume that the average duration of the sub-bands are the same, to have the same energy consumption for all the devices, i.e., $\overline{E}_i=\overline{E}_1$ for $i=1,\cdots, N_s$, we need to satisfy $\overline{P}_{\mathrm{t},i}=\overline{P}_{\mathrm{t},1}$, which can be rewritten as follows:
\begin{align}
\mu_i=\frac{r_1^{\alpha+2}}{{r_i^{\alpha+2}-r_{i-1}^{\alpha+2}}}\mu_1=\frac{\mu_1}{{i^{\frac{\alpha+2}{2}}-{(i-1)}^{\frac{\alpha+2}{2}}}},
\label{eq:mucalc}
\end{align}
where the last equality was obtained by using (\ref{eq:radius}). By using (\ref{eq:bandw1}), the bandwidth for the $i^{th}$ sub-band can be calculated as follows:
\begin{align}
W_i=\left(i^{\frac{\alpha+2}{2}}-{(i-1)}^{\frac{\alpha+2}{2}}\right)W_1.
\label{eq:fracband}
\end{align}
As we have $\sum_{i=1}^{N_s}W_i=W$, we have
\begin{align}
W=W_1\sum_{i=1}^{N_s}\left(i^{\frac{\alpha+2}{2}}-{(i-1)}^{\frac{\alpha+2}{2}}\right)=N_s^{\alpha+2}W_1
\end{align}
and by using (\ref{eq:fracband}), we have:
\begin{align}
W_i=\frac{i^{\frac{\alpha+2}{2}}-{(i-1)}^{\frac{\alpha+2}{2}}}{N_s^{\frac{\alpha+2}{2}}}W.
\end{align}

\bibliographystyle{IEEEtran}
\footnotesize
\bibliography{IEEEabrv,References}

\end{document}